\documentstyle[twocolumn,prl,aps]{revtex}

\draft
\newcommand{\be}{\begin{equation}}
\newcommand{\ee}{\end{equation}}
\newcommand{\ben}{\begin{eqnarray}}
\newcommand{\een}{\end{eqnarray}}

\newcommand{\iii}{\'{\i}}

\begin{document}
\draft
%\maketitle
\title{Werner states and the two-spinors Heisenberg anti-ferromagnet}
\author{J. Batle$^1$,  M. Casas$^1$,
 A. Plastino$^{3,\,4}$, and A. R. Plastino$^{2,\,3\,5}$}

\address {
$^1$Departament de F\iii sica, Universitat de les Illes Balears
and IMEDEA-CSIC, 07122 Palma de Mallorca, Spain \\
$^2$Faculty of Astronomy and Geophysics, National University La
Plata, C.C. 727, 1900 La Plata \\
$^3$Argentina's National Research Council (CONICET) \\
$^4$Department of Physics, National University La Plata,
C.C. 727, 1900 La Plata, Argentina\\
$^5$Department of Physics, University of Pretoria, 0002 Pretoria,
South Africa }

%\date{\today}

\maketitle

\begin{abstract}

\noindent  We ascertain, following ideas of Arnesen, Bose, and Vedral
concerning thermal entanglement [Phys. Rev. Lett. {\bf 87} 
(2001) 017901] and using the statistical tool called {\it entropic
non-triviality} [Lamberti,  Martin,  Plastino, and Rosso, Physica
A {\bf 334} (2004) 119], that there is a one to one
correspondence between (i) the mixing coefficient $x$ of a Werner
state, on the one hand, and (ii) the temperature $T$ of the
one-dimensional Heisenberg two-spin chain with a magnetic field
$B$ along the $z-$axis, on the other one.  This is true  for each
value of $B$ below a certain critical value $B_c$. The pertinent
mapping depends on the particular $B-$value one selects within
such a range.

\noindent
 Pacs: 03.67.-a; 89.70.+c; 03.65.-w; 02.50.-r

\end{abstract}
%\vspace{.5cm}

%\vskip 5mm \noindent \hskip 2cm Keywords: Quantum Entanglement;
%classical entanglement; Quantum Information Theory
%\newpage
\section{\bf Introduction}

Entanglement is one of the most fundamental issues of quantum
theory \cite{L1,L2,L3,L4,L5,L6} and the so-called Werner states
\cite{We89} have played a distinguished role in the unravelling of
the fascinating issues at play (see, for instance,
\cite{Pop1,Pop,Bennett}). The Werner density matrix reads \be
\label{uno} \rho_W\,=\,x|\Phi^{+}\rangle\langle
\Phi^{+}|\,+\,\frac{1-x}{4}I,\ee where $|\Phi^{+}\rangle$ is a
Bell state (maximally entangled). The state (\ref{uno}) is
separable (unentangled) for the mixing coefficient $x\le 1/3$
\cite{We89}. For $x > 1/3$ they are entangled and violate
 the CHSH inequality for $x > 1/\sqrt 2$ \cite{Pop1,Pop,Bennett}.
 %%%%Below modification for Referee's Comment 1) %%%%%
We see that  Werner states are mixtures of noise and a maximally
entangled state, and therefore, for  values of the mixing
parameter $x >1/3$ they are entangled, violate the CHSH inequality
and exhibit nonclassical
 features \cite{Pop1,Pop,Bennett}.
%%%%%%%%%%%%%%%%%%%%%%%%%%%%%%%%% end of 1) %%%%%%%%%%%%

%\subsection{The Heisenberg model}

\noindent Following the interesting work of Arnesen {\it et al.}
\cite{Arnesen}, we concern ourselves with the issue of {\it
thermal entanglement} and consider the Hamiltonian for the 1D
Heisenberg spin chain with a magnetic  field of intensity $B$
along the $z$-axis

\begin{equation}\label{hamiltonian}
H\,=\,\sum_{i=1}^{N} (B\sigma^{i}_{z} \, + \, J_H \vec \sigma^{i} \,
\vec \sigma^{i+1}),
\end{equation}
\noindent where $\sigma^{i}_{x,y,z}$ stand for the Pauli
matrices associated with spin $i$ and periodic boundary conditions are 
imposed ($\sigma^{N+1}_{\mu}=\sigma^{1}_{\mu}$). $J_H$ gives the strength of
the spin-spin repulsive interaction (only the anti-ferromagnetic
($J_H>0$) instance is discussed). If we limit ourselves to the case
$N=2$, we will be dealing with two spinors, i.e., with a
two-qubits system. For ``thermal equilibrium"  we should consider
\cite{Arnesen} the thermal state \be \label{rhot}
\rho(T)=\frac{\exp(-\frac{H}{k_{B}T})}{Z(T)},\ee with $Z(T)$  the
partition function. Expressing both $H$ and $\rho(T)$ in the
computational basis $|00\rangle,|01\rangle,|10\rangle,|11\rangle$
we obtain

\begin{equation}
H = \left( \begin{array}{cccc}
2J_H+2B & 0 & 0 & 0\\
0 & -2J_H & 4J_H & 0\\
0 & 4J_H & -2J_H & 0\\
0 & 0 & 0 & 2J_H-2B \end{array} \right).
\end{equation}
After defining, for convenience's sake, \newline \noindent
$e_{wmy}=\exp{\,(-2w-2y)};$
\newline \noindent $e_{wp}=\exp{\,(-2w)}+\exp{\,(6w)};$\newline \noindent
$e_{wm}=\exp{\,(-2w)}-\exp{\,(6w)}$; \newline \noindent $e_{wpy}=
\exp{\,(-2w+2y)},$ \newline \noindent with $w=J_H/k_{B}T$ and
$y=B/k_{B}T$, we also get

\begin{equation}
\rho(T) = \frac{1}{Z(T)}\left( \begin{array}{cccc}
e_{wmy} & 0 & 0 & 0\\
0 & e_{wp}/2 & e_{wm}/2 & 0\\
0 & e_{wm}/2  & e_{wp}/2 & 0\\
0 & 0 & 0 & e_{wpy} \end{array} \right),
\end{equation}
\noindent  The concurrence of $\rho(T)$ reads \cite{Arnesen} \ben
C&=&0;\,\,\,
\,\,\,\,\,\,\,\,\,\,\,\,\,\,\,\,\,\,\,\,\, \,\,\,\,\,\,\,\,\,\,
\,\,\,\,\,\rm{for}\,\,\, T \ge T_c, \cr C&=&
\frac{e^{8w}-3}{1+e^{-2y}+e^{2y}+e^{8w}}; \,\,\rm{for}\, T <
T_c,\een  For our purposes we must emphasize that there is no
entanglement beyond a certain critical temperature $T_c=8J_H/(k_B\ln
3)$ \cite{Arnesen}. Clearly,  the temperature plays a sort of
``mixedness-role", since it is well-known that entanglement
vanishes for a high enough degree of mixing.
%For the Hamiltonian we are
%consider here
%\lq\lq thermal entanglement" decreases as we increase $T$.
%%%%%%%%%%%%%%Comment 4)%%%%%%%%%%%%%%%%%%%
Notice that $T_c$ is independent of $B$ and thus an intrinsic
structural property that depends (linearly) just
 on the interaction strength $J_H$. For strong enough coupling strengths
 we encounter non null entanglement for a wide range of temperature values.
 %%%%%%% Comment 4)%%%%%%%%%%%%%%%%

%%%%%% End of 4) %%%%%%%%%%%%%%%%  =20
\noindent Also, there is a change in the structure of the ground state of
hamiltonian (\ref{hamiltonian}) when the magnetic field reaches
the critical value $B_c=4J_H$.

\noindent  Remarkably, the Werner states (\ref{uno}) and the thermal
states (\ref{rhot}) of the 1D Heisenberg model for $N=2$ can be
related by a one to one correspondence between the mixing
parameter $x$ and the temperature $T$ for each value of $B \le
B_c$. By means of this relation we explore for both states the
evolution of the entanglement of formation as a function of the
temperature.

The paper is organized as follows: In Section II we introduce the 
entropic non triviality measure based on the Jensen-Shannon divergence, 
which is the basic tool that provides a deeper insight into the behaviour 
of the entanglement of
formation as a function of $T$ around $B_c$.
 Our results are reported in
Section III, where we provide the exact mapping
between Werner states and the aforementioned thermal states.
Finally some conclusions are drawn in Section IV.

\section{ The Jensen-Shannon divergence and the entropic non-triviality
measure}

We review now a statistical information measure with which
we will be concerned in the present work. 
Let ${\vec P}_{(k)} \in  \Omega \subset{\cal R}^{N}$, with $k=1,2$, denote
two different probability distributions for a particular set of
$N$ accessible micro-states.  The components of the two
probability vectors ${\vec P}_{(k)}$ must satisfy the following
two constraints: {\it a)\/} $\sum_{j=1}^N~p_j^{(k)}=1,$ and {\it
b)\/} $0 \leq p_j^{(k)} \leq 1~\forall j$. The set $\Omega$
defined by these constraints is the simplex $S_N$, which is a
convex $(N-1)$-dimensional subset of $R^N$. A quite important,
information-theoretical based divergence measure between ${\vec
P}_{(1)}$ and ${\vec P}_{(2)}$ was originally introduced by Rao
\cite{rao} and used  by several authors \cite{lin}. It came
afterwards to be called the Jensen-Shannon divergence (JSD)
\cite{pregiven,lamberti,topsoe} that
\begin{itemize} \item induces a {\it true} metric in $\Omega
\subset{\cal R}^{N}$, being indeed the square of a metric
\cite{topsoe}, and \item is intimately related to the
Kullback-Leibler relative entropy  $K$  for two probability
distributions ${\vec P}_{(1)}$ and ${\vec P}_{(2)}$, given by
\cite{Kullback51}
\begin{equation}
K[{\vec P}_{(1)} \vert {\vec P}_{(2)}]~= ~\sum_j~p^{(1)}_j~\ln
\left(~p^{(1)}_j~/~p^{(2)}_j~\right). \label{KL-entropy}
\end{equation}
\end{itemize}
We first define
\begin{equation}
\label{uno1} J_0[ {\vec P}_{(1)}, {\vec P}_{(2)} ]=
K[ {\vec P}_{(1)} \vert ( {\frac{1}{2}}{\vec
P}_{(1)}+{\frac{1}{2}}{\vec P}_{(2)}) ],
\end{equation}
and then the symmetric quantity
\begin{eqnarray}
\label{dos2} J_1[{\vec P}_{(1)}, {\vec P}_{(2)}]&=&
J_0[{\vec P}_{(1)}, {\vec P}_{(2)} ]+  J_0[{\vec
P}_{(2)}, {\vec P}_{(1)} ]  \\ \nonumber &=& 2S[
\frac{1}{2} {\vec P}_{(1)}+
              \frac{1}{2} {\vec P}_{(2)} ] \cr
   &-& S[ {\vec P}_{(1)} ]
   - S[ {\vec P}_{(2)} ],
\end{eqnarray}
where $S =-\sum_j p_j \ln p_j $ is the Shannon logarithmic information measure.
Let now $\pi_1,~\pi_2 > 0$; $\pi_1+\pi_2=1$ be the ``weights" of,
respectively, the probability distributions ${\vec P}_{(1)},~{\vec
P}_{(2)}$. The JSD reads \ben \label{tres3}
J^{\pi_1,\pi_2}\left[{\vec P}_{(1)}, {\vec P}_{(2)} \right]&=&
S\left[ \pi_1 {\vec P}_{(1)} +
        \pi_2 {\vec P}_{(2)} \right] \cr
   &-& \pi_1S\left[ {\vec P}_{(1)} \right]
   - \pi_2S\left[ {\vec P}_{(2)} \right],
\een
 which is a positive-definite quantity that vanishes iff ${\vec
P}_{(1)} = {\vec P}_{(2)}$ almost everywhere
\cite{pregiven,lamberti}. In the particular case {\it to be used
in this work}, $\pi_1=\pi_2= 1/2 $, the measure (\ref{tres3}) is
symmetric. Notice also that
$J^{\frac{1}{2},\frac{1}{2}}=\frac{1}{2}J_1$.

%\vskip 3mm \noindent {\small{The entropic non-triviality measure }}
%\vskip 3mm

Another statistical tool that we need is the so-called entropic
non-triviality measure \cite{nontr}. The statistical
characterization of {\it deterministic\/} sources of {\it
apparent\/} randomness performed by many authors during the last
decades has shed much light into the intricacies of dynamical
behavior by describing the unpredictability of dynamical systems
using such tools as metric entropy, Lyapunov exponents, and
fractal dimension \cite{beck}. It is thus possible to {\it i)\/}
detect the presence and {\it ii)\/} quantify the degree of
deterministic chaotic behavior. Ascertaining the degree of
unpredictability and randomness of a system is {\it not
automatically tantamount to adequately grasp the correlational
structures that may be present}, i.e., to be in a position to
capture the relationship between the components of the physical
system. %%%%%%%%Comment 3) : frase eliminada %%%%%%
  Certainly, the opposite extremes of {\it i)\/} perfect
order and {\it ii)\/} maximal randomness possess no structure to
speak of. In between these two special instances a wide range of
possible degrees of physical structure exists, degrees that should
be reflected in the features of the underlying probability
distribution (PD). One would like that they be adequately captured
by some functional of PD in the same fashion that Shannon's
information measure captures randomness. A candidate to this
effect has come to be called the \textsc{entropic non-triviality}
(also, ``statistical complexity") $\mathcal{C}$ \cite{nontr}, that
should, of course, vanish in the two special extreme instances
mentioned above.

%%%%%%%%%%%%%%%%%%%%%%%%%%%%%%%%%%%%New text for Referee's  comment 2) =
%%%%
In order to illustrate these assertions we  can refer  to the
celebrated logistic map  $F: x_n \rightarrow x_{n+1}$
\cite{COPING}, where one focuses attention upon the ecologically
motivated, dissipative system described by the first order
difference equation
\begin{equation}
x_{n+1}=r~x_n ~(1- x_n)~\quad  (~0 \le x_n \le 1~, ~0< r \le 4~)
\label{j1}
\end{equation}
whose dynamical behavior is controlled by $r$. For values of the
control parameter $1 < r < 3$ there exists only a single
steady-state solution. Increasing the control parameter past $r=3$
forces the system to undergo a period-doubling bifurcation. Cycles
of period $8,~16,~32,~\cdots$ occur and, if $r_n$ denotes the
value of $r$ where a $2^n$ cycle first appears, the $r_n$ converge
to a limiting value $r_\infty\cong 3.57$ \cite{COPING}. As $r$
grows still more, a quite rich, and well-known structure becomes
apparent. A cascade of  period-doubling  occurs as $r$ increases,
until, at $r_\infty$, the maps becomes chaotic and the attractor
changes from a finite to an infinite set of points. For $r >
r_\infty$ the orbit-diagram reveals an ``strange" mixture of order
and chaos, with notable windows of periodicity  beginning near
$r=3.83$. As shown in \cite{lamberti}, the measure  $\mathcal{C}$
is able de give detailed account of the pertinent dynamical
features (see, for instance, Figs. 2-4 of  \cite{lamberti}).
%%%%%%%%%%%%%%%%%%%%%%%%%%%%%%%%%%%%%% end of 2) %%%%%%%%%%%%%%%%

Typically, the measure  $\mathcal{C}$ is the product of two
quantities: (1) a normalized entropy (i.e., whose values range
between 0 and 1) $H=S/S_{max}$ (the denominator is the largest
possible value for $S$) and (2) a chosen ``distance" $d(\vec
P,\vec P_u )$ in probability space that measures ``how far" (in
this space) the actual PD $\vec P$ lies from the {\it uniform} PD
(of maximal entropy)  $\vec P_u$, i.e., \be \label{defi}
{\mathcal{C}} =  H d(\vec{P}, \vec{P_u})\ee (see, for example,
\cite{nontr,sc1,sc4,sc2,sc3,sc5,sc6} and references therein).  In
the present work we are concerned with the probability set
 (density matrix' eigenvalues)
 associated to any state $\rho$ and then, with reference to (\ref{defi}),
  take  as $H$ the normalized
von Neumann entropy $H_{vN}$ of the relevant state $\rho$ and as
$d$ the Jensen-Shannon divergence that will here measure the
``distance" from the relevant state to the maximally mixed one
$\rho_{MM}$ (proportional to the identity matrix $I$). Our
entropic non-triviality acquires then the aspect (Cf. Eq.
(\ref{tres3})) \be C_{JS}(\rho)=J^{1/2,1/2}(\rho,\rho_{MM})
H_{vN}( \rho).\ee

\noindent One may wonder whether it is possible to built up an entropic 
non-triviality measure that would depend not on density matrices but 
directly on wave functions.
The concomitant extension is of a rather non trivial nature, but can be 
accomplished and steps toward its implementation are currently being 
undertaken. They will be published in the near future.
\section {\bf Results}

Our main interest refers to the ``temperature-evolution" of both
$C_{JS}$ and the entanglement of formation $E_f$ as  $T$
diminishes, and, eventually, in the limit case $T \rightarrow 0$,
where a change in the structure of the ground state of hamiltonian
(\ref{hamiltonian}) is detected for $B_c=4J_H$. For convenience,
we take $J_H=1$ from now on. In the limit $T \rightarrow 0$ one
can state that

\begin{itemize}
        \item 1) [$B<B_c$, $T=0$] the ground state (gs)
        is non-degenerate and equal to the singlet state
($E_f=1$, $C_{JS}=0$ for all $B$)
\begin{equation} \rho_{1}^{gs}(T=0)=\left( \begin{array}{cccc}
0 & 0 & 0 & 0\\ 0 & \frac{1}{2} & -\frac{1}{2} & 0\\ 0 &
-\frac{1}{2} & \frac{1}{2} & 0\\ 0 & 0 & 0 & 0 \end{array}
\right),
\end{equation}

        \item 2) [$B=B_c$, $T=0$] the gs state becomes two-fold 
degenerate,
        the corresponding two eigenstates being the singlet state 
{\it and}
$|11\rangle$ ($E_f=E_f^c$, $C_{JS}=C_{JS}^{c}$)

\begin{equation}
\rho_{2}^{gs}(T=0) =\left( \begin{array}{cccc}
0 & 0 & 0 & 0\\
0 & \frac{1}{4} & -\frac{1}{4} & 0\\
0 & -\frac{1}{4} & \frac{1}{4} & 0\\
0 & 0 & 0 & \frac{1}{2} \end{array} \right),
\end{equation}

        \item 3) [$B>B_c$, $T=0$] the ground state is now 
$|11\rangle$, which
has no entanglement ($E_f=0$, $C_{JS}=0$ for all $B$)

\begin{equation}
%\ben
\rho_{3}^{gs}(T=0) = |11\rangle\langle 11| =  \left(\begin{array}{cccc}
0 & 0 & 0 & 0\\
0 & 0 & 0 & 0\\
0 & 0 & 0 & 0\\
0 & 0 & 0 & 1 \end{array} \right).
\end{equation}
%\end{eqnarray}

\end{itemize}
It is clear that, for fixed $J_H$ and at $T=0$, the entropic
no-triviality is zero for all $B$ except for $B_c$, which is
tantamount to asserting that $C_{JS}$ {\it ``detects" the
``critical" magnetic strength at which the gs-structure changes.}
%%first result

\vskip 3mm \noindent We return attention now to the Werner state \be
 \label{We89} \rho_W\,=\,x|\Phi^{+}\rangle\langle \Phi^{+}|\,+\,(1-x)
 \rho_{MM},\ee
where $|\Phi^{+}\rangle$ is a  (maximally entangled) Bell state.
This state is unentangled for $x\le\frac{1}{3}$ \cite{We89}. Fig.
1 depicts  $C_{JS}(\rho)$ vs. $T$ and $E_f(\rho)$ vs. $T$ for the
1D Heisenberg model in the case $B=0$. For fixed values of $J_H$
and $B$, there exists a critical temperature
$T_c=\frac{8J_H}{k_B\ln 3}$ above which the system is no longer
entangled. The same quantities are depicted in the inset for the
Werner state $\rho_W$. If one considers the plot $C_{JS}(\rho_W)$
vs. $E_f(\rho_W)$ and its thermal  Heisenberg counterpart
$C_{JS}(\rho)$ vs. $E_f(\rho)$ for $B=0$, the two graphs coincide.
In point of fact, there is a mapping between both states, if we
take 
\be \label{map} 
x =\frac{2}{3}\left[\frac{e^{8\omega}-3}{1+e^{-2y}+e^{2y}+
e^{8\omega}}+\frac{1}{2}\right], 
\ee  
there is a one to one
correspondence between the mixed parameter $x$ of the Werner
states and the temperature $T$ of the states (\ref{rhot}) for each
value of $B \le B_c$ (see Fig. 2) so that $x$  can be regarded as
an effective temperature $T_{eff}$. We remark that
\begin{enumerate} \item at $x=1$ ($T=0$)
both states are maximally entangled (and have zero complexity)
until a critical point is reached,  \item at the critical point
$x=\frac{1}{3}\,\,\, \rightarrow \,\,\,T_c=\frac{8J_H}{k_B\ln 3}$
the two states have the same $C_{JS}$. Notice that all the curves
of Fig. 2 (for any several $B-$values)  intersect at the point
$(x=1/3,1/k_B T_c)$,
\item when $x=0$
($T \rightarrow \infty$) both states are unentangled (also with
$C_{JS}=0$).
\end{enumerate}
\noindent In the vicinity of $B_c$, the plots $C_{JS}$ vs. $E_f$ (as
parameterized by $T$) provide a good insight into the ensuing
transition mechanism. In Fig. 3 we plot $C_{JS}(\rho)$ vs.
$E_f(\rho)$ for several $B-$values  near the critical point
$B_c=4$. Note that, as we approach the critical point from either
the left (inset) or the right, the entropic non-triviality
augments.
 Pay special attention to the rightwards ($B_{c}^{+}$) approach.
 We not only appreciate a maximum in $C_{JS}$, but notice also that
$E_f$ is optimal at $B=4.1$.
 Thus,  maximum $C_{JS}$ equals maximum $E_{f}$ at the critical
point. Vertical and horizontal dashed lines correspond to critical
values at $T=0, B=B_c$.
As seen from $\rho^{gs}$, there is sudden
entanglement-change as $B$  crosses $B=B_c$. As $C_{JS}$ can only
detect changes in state-structure, its variations can only be
produced gs-transitions. Its ``transition-detection" feature
improves its accuracy in the limit $T \rightarrow 0$.

Transition details near $B_c=4$ are examined in an even closer
fashion in Fig. 4,  a $C_{JS}$ vs. $E_f$ plot. We depict things
for $B_{c}^{+}=4.001$ (solid line) and for $B_{c}^{-}=3.99$
(dot-dashed line). Zone I is the region with $E_f > E_f^c$
($B_{c}^{-}$), wherefrom, as we increase $T$, the state loses
entanglement while increasing its entropic non-triviality
(dot-dashed line). See in the inset (solid thick line) the
crossing at $E_f^c$, enhanced in order to appreciate details of
the evolution towards maximal entanglement and maximal complexity
in the limit $T \rightarrow 0$. On the other hand, if the initial
point occurs at $B_{c}^{+}$, no matter whether  the temperature
changes start at either $T=0$ or $T=\infty$, we end up at the same
$E_f=0$ point. In particular, for $B=B_c$, there is a single curve
that connects the $T=0$ critical point with the $T \rightarrow
\infty$ limit (where $C_{JS}=E_f=0$). The double-arrow connecting
the curves in the two zones illustrates the fact that a path
beginning in Zone I (dot-dashed line) cannot smoothly merge with
the curve depicted in Zone II (solid line),  appearances
notwithstanding. A glimpse at the inset clarifies the situation.
Vertical and horizontal dashed lines correspond to critical values
at $T=0, B=B_c$.

Fig. 5 is a $C_{JS}$ vs. $E_f$ plot that intends to depict things
in the immediate vicinity of the critical point $B_c=4$. As in
Fig. 4, here we fix a very low $T$ (of the order $10^{-3}$) while
varying $B$. The solid line corresponds to the entropic
non-triviality. See the quasi-circular ``motion"  around $B_c$ as
we decrease the temperature. Inset a) depicts $C_{JS}$ vs. $E_f$,
which, surprisingly, coincides with the ``envelope-curve" (Zone I
+ Zone II) of Fig. 4. This remarkable fact can be explained
because in Fig. 4 both the ``upper" branch of the curves in Zone I
{\it and} the curve of Zone II are drawn for low $T$-values nearby
$B=Bc$. Since here we ``sit" at the critical point (and very low
temperatures are used) both curves of inset a) look like a
continuous lines. In other words, only a low $B-$range of values
around $B_c$ is relevant at such low temperatures. Inset b)
depicts the behavior of the degree of mixture $R=1/Tr[\rho(T)^2]$
vs. $E_f$. In the limit $T \rightarrow 0$, $R$
 reaches $R=2$, as expected. Vertical and
horizontal dashed lines correspond to critical values at $T=0,
B=B_c$.

%\end{itemize}

\section{Conclusions}
In this paper we have shown that the Jensen-Shannon entropic
non-triviality measure $C_{JS}$  is able to detect the ``critical"
magnetic strength at which the gs-structure changes. Also, via
$C_{JS}$ we discover that there is a one to one correspondence
between Werner states and the Heisenberg anti-ferromagnet for
values of $B \le B_c$.
%The mapping reads \be \label{map} x =
%\frac{2}{3}\left[\frac{\exp{8\omega}-3}{1+\exp{-2y}+\exp{2y}+
%\exp{8\omega}}+\frac{1}{2}\right]. \ee
  As
shown in Fig. 2, it is possible to assign to the mixing Werner
parameter $x$ an effective temperature $T_{eff}$ for each value of
$B \le B_c$. Finally, the physics in the immediate vicinity of the
critical point $B_c=4$ can be analyzed in detailed fashion using
$C_{JS}$. In the one dimensional Heisenberg model, for two
spinors, maximum entanglement implies maximum entropic
non-triviality in the vicinity of the corresponding critical
point.

\vskip 3mm
{\bf Acknowledgements}
\vskip 3mm
This work was partially supported by the MEC grant BFM2002-03241 (Spain)
and FEDER (EU), by the Government of Balearic Islands and by CONICET
(Argentine Agency).

\vskip 3mm
FIGURE CAPTIONS
\vskip 3mm

\noindent Fig. 1- Evolution of (1) the entanglement of formation
$E_f$ and (2) the Jensen-Shannon complexity $C_{JS}$ vs. the
temperature $T$ for the thermal state $\rho(T)$ of two-qubits in
the 1D Heisenberg model. For given values of $J_H$ and $B$ there
exists a critical temperature $T_c=\frac{8J_H}{k_B\ln 3}$ above
which the system is no longer entangled. The same quantities are
depicted in the inset for the Werner state $\rho_W$. There is a
one to one mapping between both type of states for $B \le B_c$. See text =
for
details.

\vskip 0.5cm \noindent Fig. 2- Mapping of the mixed Werner
parameter $x$ as a function of $1/T$ for several values of $B$.
The horizontal lines correspond to $x=1/3$ and $x=2/3$,
respectively.

\vskip 0.5cm

\noindent Fig. 3- Plot of the complexity $C_{JS}$ vs. the
entanglement of formation $E_f$ near the critical point $B_c=4$.
As we approach the critical point from either left (inset) or
right, the complexity augments. In the $B_{c}^{+}$ case we not
only have a maximal $C_{JS}$, but also an optimal $E_f$. Vertical
and horizontal dashed lines correspond to critical values at $T=0,
B=B_c$. See text for details.

\vskip 0.5cm

\noindent Fig. 4- Same as before, but with additional details. We
have $B_{c}^{+}=4.001$ (solid line) and $B_{c}^{-}=3.99$
(dot-dashed line). Zone I (II) denotes the region with $E_f >
E_f^c$ and $B_{c}^{-}$ ($E_f < E_f^c$ and $B_{c}^{+}$). In the
inset the crossing region at $E_f^c$ is enhanced in order to see
how details of the evolution around maximum entanglement and
maximum complexity in the limit $T \rightarrow 0$. See text for
details.

\vskip 0.5cm

\noindent Fig. 5- Plot of the complexity $C_{JS}$ in the vicinity
of the critical point $B_c=4$. The inset a) depicts $C_{JS}$ vs.
$E_f$, which, surprisingly, coincides with the enveloping curve
(Zone I + Zone II) of Fig. 3. Inset b) depicts the behaviour of
the degree of mixture $R=1/Tr[\rho(T)^2]$ vs. $E_f$. In the limit
$T \rightarrow 0$, $R$ reaches the value $R=2$, as expected.
Vertical and horizontal dashed lines correspond to the critical
values at $T=0, B=B_c$. See text for details.

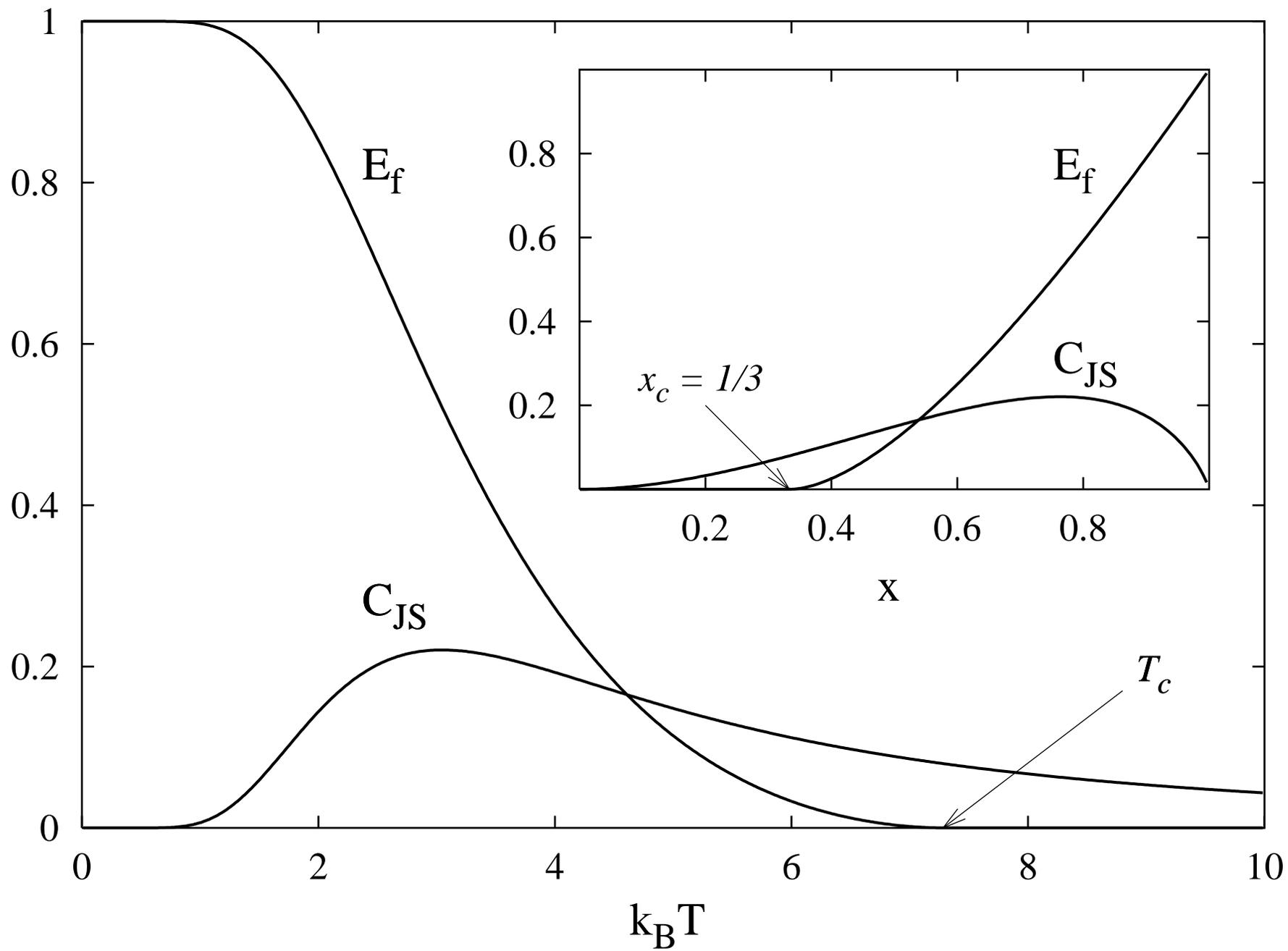
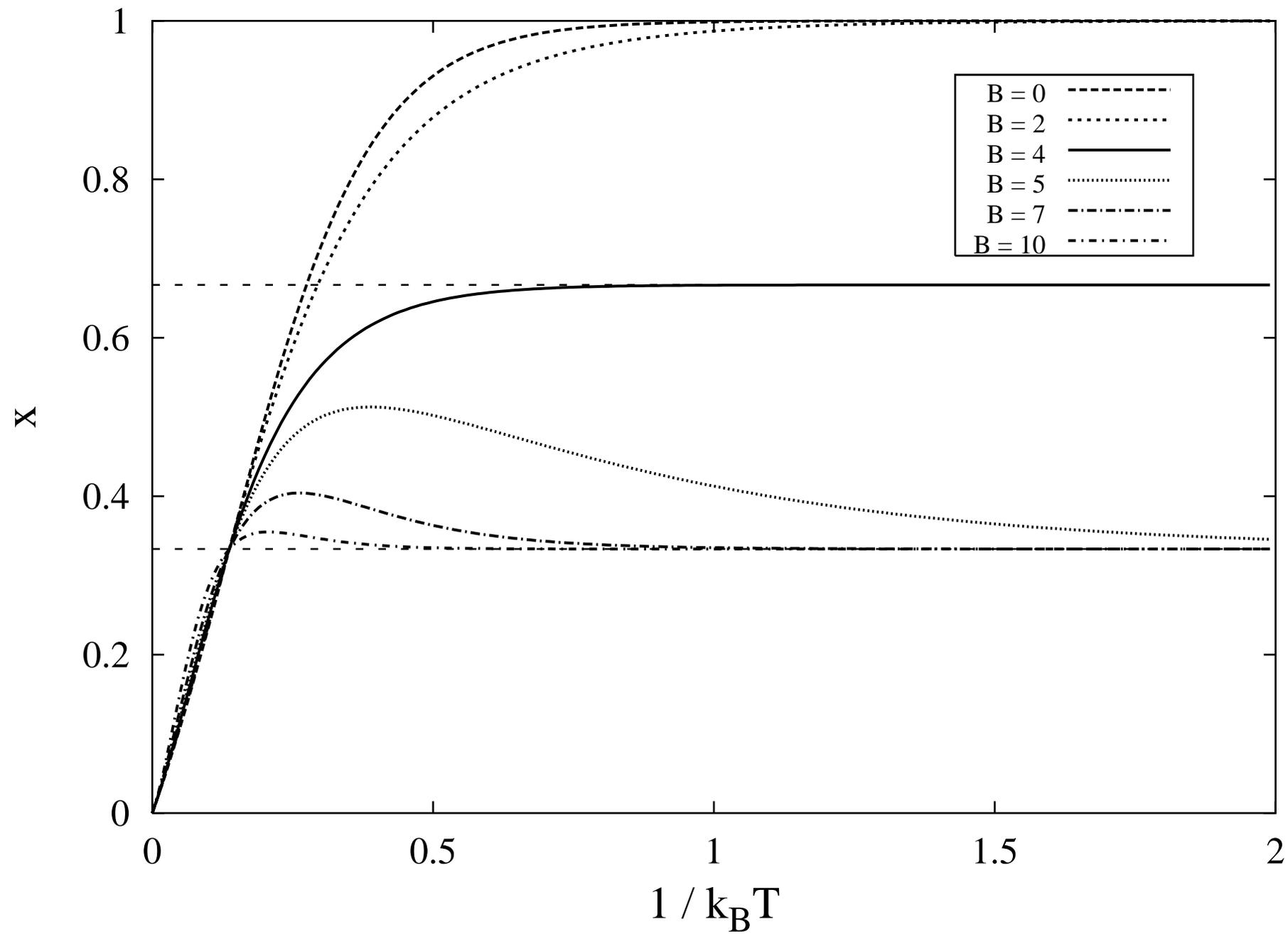
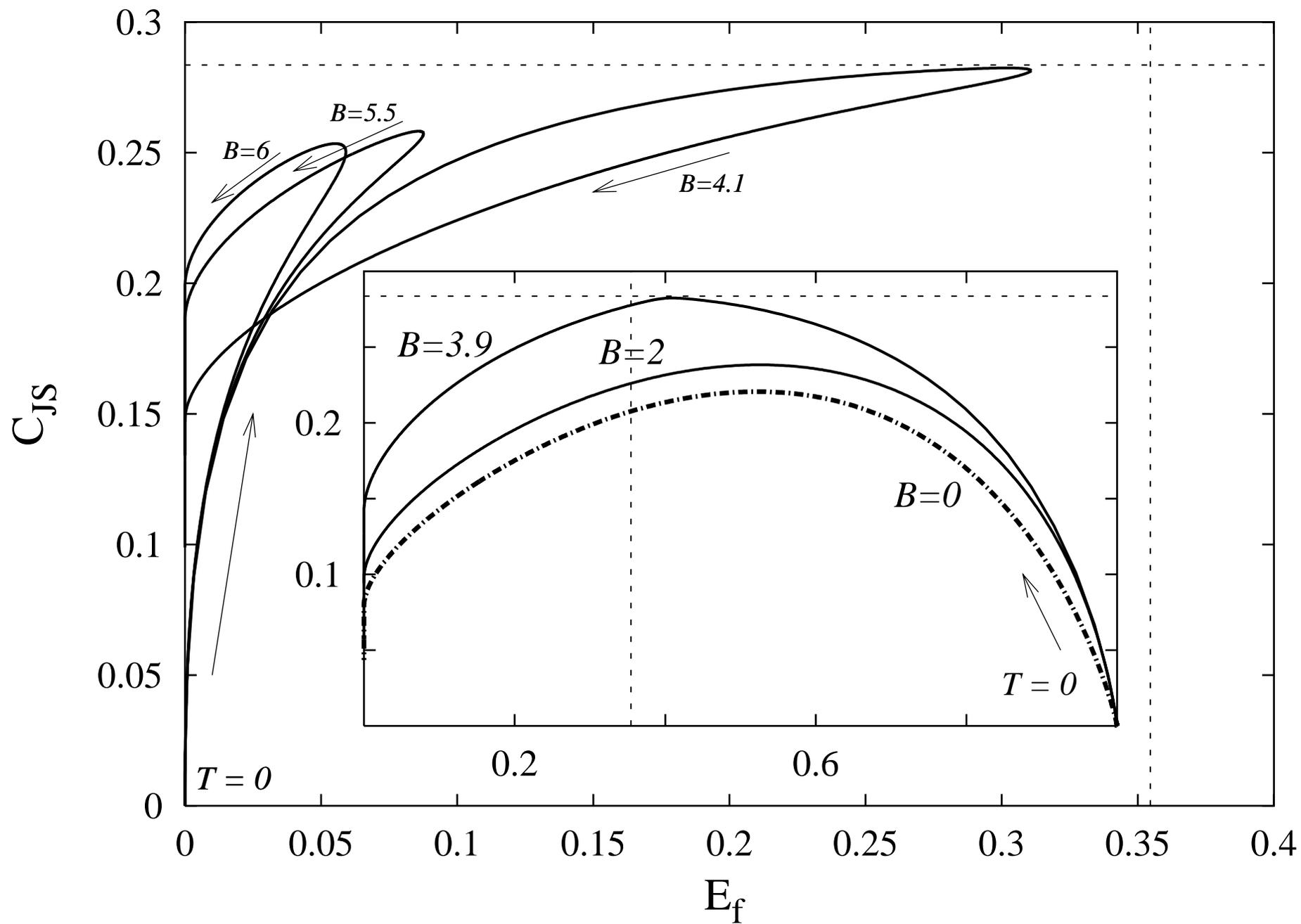
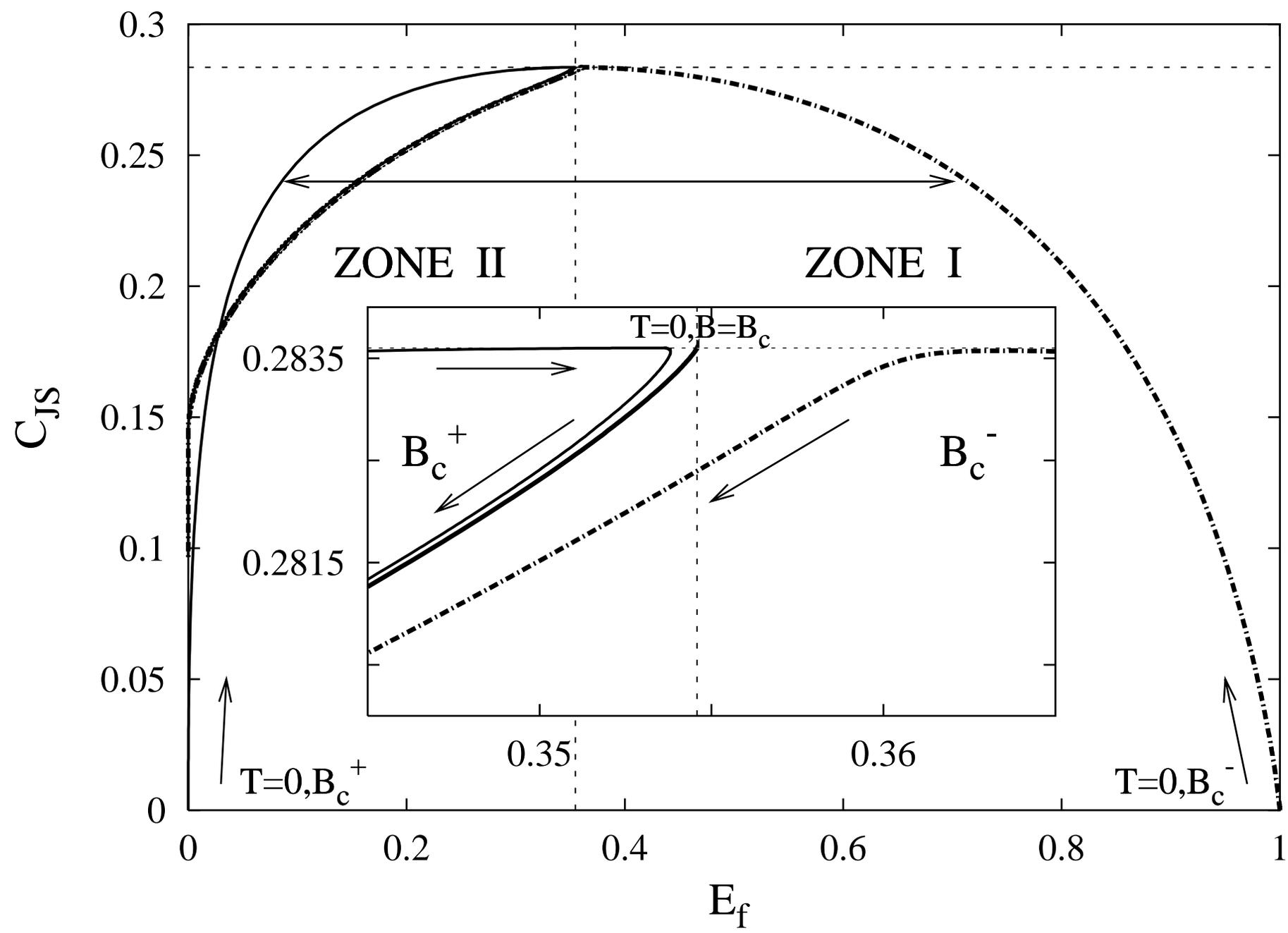
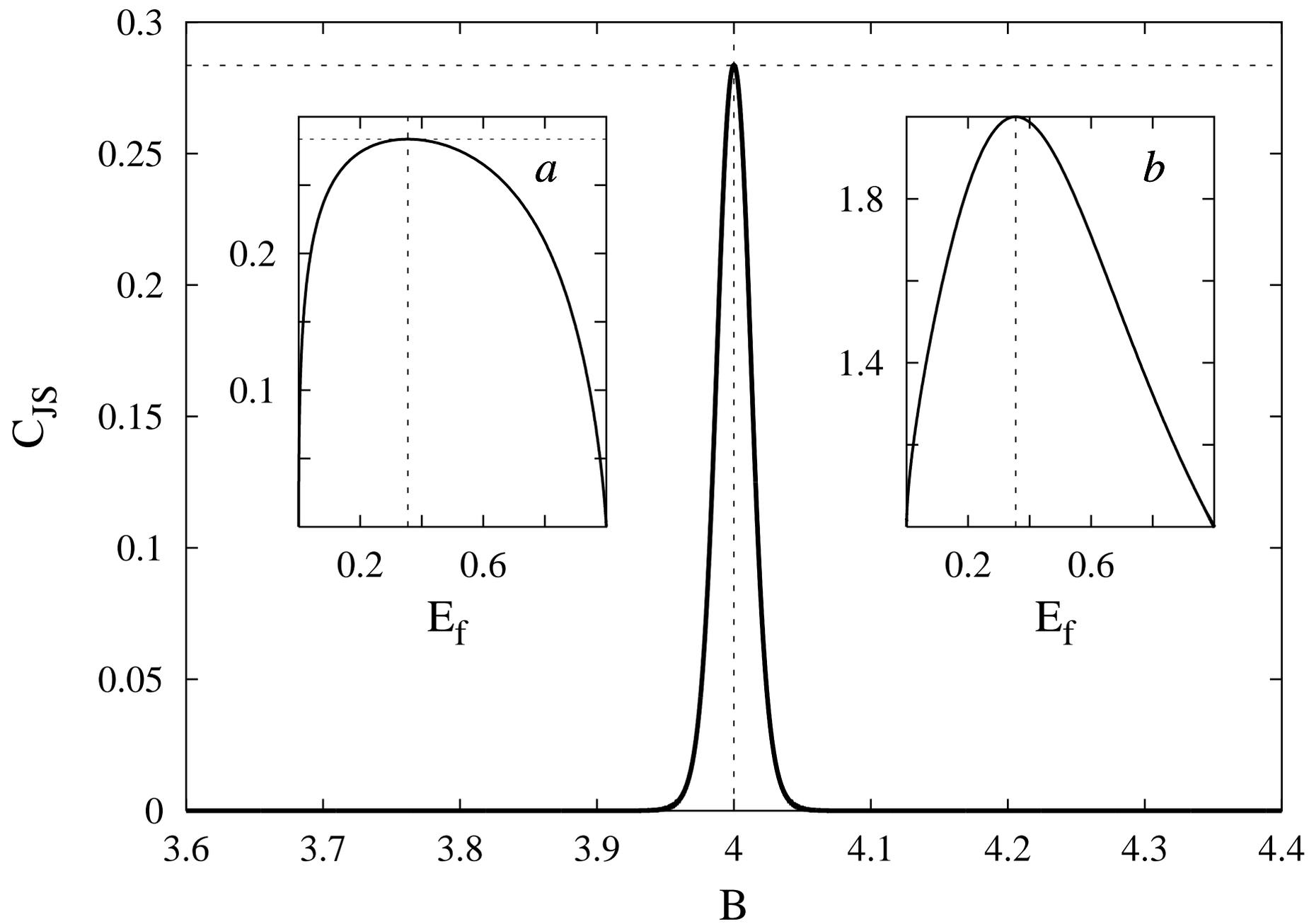
\end{document}